\documentclass[12pt, a4paper]{article}
\usepackage[font=footnotesize]{caption}
\usepackage{graphicx}
\usepackage{authblk}
\usepackage{booktabs}
\usepackage{verbatim}
\usepackage{url}
\usepackage{caption}
\usepackage{subcaption}
\usepackage{indentfirst}
\usepackage{url}
\usepackage{amsmath}
\author{Alexandre Benatti$^1$ and Luciano da F. Costa$^1$}

\date{%
    $^1$S\~ao Carlos Institute of Physics, \\
    University of S\~ao Paulo, \\
    S\~ao Carlos, SP, Brazil.\\[2ex]
    15th Jun.~2023
}

\begin{document}

\title{On the Transient and Equilibrium Features \\
of Growing Fractal Complex Networks}

\maketitle

\begin{abstract}
Complex networks have certain properties that distinguish them from their respective uniform or regular counterparts. One of these properties is the variation of topological properties along different hierarchical levels. In this work, we study how networks that are constructed by repeatedly incorporating a given motif exhibit this property. A motif is henceforth understood as a small subgraph with a reference node where the incorporation respectively occurs. We generate fractal networks using different motifs and observe how their topology changes depending along the growth stages. Two regimes are respectively identified: transient and equilibrium. The former is characterized by significant topological changes that depend on the motif topology, while the equilibrium regime shows more stable, parallel trajectories. A more systematic analysis revealed that the betweenness centrality and the average shortest path lengths were the main topological properties that change along the network growth.
\end{abstract}

\section{\label{sec1}Introduction}
The intrinsic ability of graphs and complex networks to represent virtually any discrete structure, abstract or from the real world, has motivated many works aimed at characterizing and modeling complex systems in terms of networks (e.g.~\cite{barabasi2013network,newman2018networks,costa2011analyzing}).

Typically, network complexity has been associated with topological heterogeneities, in the sense of more complex networks not being well described in terms of simple measurements, such as the node degree (e.g.~\cite{da2018complex}). In contrast, predominantly homogeneous, or even regular, networks have been understood as being relatively simpler.

Given that several abstract and real-world systems and structures have an underlying hierarchical organization, one important aspect of the complexity of networks relates to how the respective topological properties change along successive hierarchies. In principle, complex networks with heterogeneities extending along all, or several, hierarchies would be understood as being particularly intricate.

While \emph{trees} correspond to the most essentially hierarchical networks, there are other types of networks presenting an intrinsic hierarchical organization, including the \emph{fractal networks} (e.g.~\cite{song2005self,barabasi2001deterministic,dorogovtsev2002pseudofractal}), characterized by the recurrent repetition of a basic pattern along several topological scales or hierarchies.

The concept of fractal networks rises an interesting problem regarding the characterization of their topological properties and complexity across different hierarchies. Two points are especially relevant from the perspective of current work: (a) understanding how the topological properties of the networks change across successive hierarchies and (b) identifying the topological properties that undergo the most significant changes across these hierarchies. In this way, a fractal network with multiple topological properties that change across multiple hierarchies can be considered more complex.

The box-covering method~\cite{song2005self} provides an interesting approach to the concept of fractal complex networks. This method consists of dividing the network into boxes of different sizes and counting the number of boxes needed to cover the entire network. If the network is fractal, then the relationship between the number of boxes and the box size follows a power law, and the exponent of this power law is called the fractal dimension. The fractal dimension measures how the network fills space, reflecting its degree of self-similarity and heterogeneity. 

The current work aims at exploring the concepts and questions mentioned above. More specifically, we start by defining a type of fractal network in which a basic pattern, or motif, is incorporated recurrently during the network growth. In addition, the motifs are understood to have one or more reference nodes, which define the positions where the recurrent incorporation of the motif takes place. 

In particular, we consider networks derived from nine specific motifs. We measure four key topological properties (clustering coefficient, betweenness centrality, shortest path, and hierarchical degree) at each stage of network growth, which allows us to quantify how the topology changes as more hierarchies are added. The sequence of networks obtained at each stage thus forms a \emph{trajectory} in the four-dimensional feature space.

By generating fractal networks and obtaining the respective trajectories, several interesting results have been identified. First, all obtained trajectories that have been found can be divided into an initial \emph{transient} regime, followed by respective \emph{equilibrium} dynamics. Of particular interest, all trajectories are mostly parallel in the equilibrium phase.  

Regarding the topological changes undergone by the networks along the transient regime, it has been found that greater modifications can be expected for different initial motifs. The fact that all trajectories have similar orientations during the equilibrium regime allowed the identification of the main topological changes which undergo most of the changes in this regime. Interestingly, only half of the considered topological features have been found to be more intensely modified along the incorporation of successive hierarchies.

Our work is structured as follows. Section~\ref{sec:related} presents the main related works, including those proposing fractal models, also providing an overview of some applications of this type of network. In Section~\ref{sec:methodology}, we describe the concepts and methods used in this study, including a \emph{fractal generator} model, and explain the topological measurements adopted to evaluate the growth progress of fractal networks. Finally, the results and analysis are described in Section~\ref{sec:results}.

\section{Related works} \label{sec:related}

In this section, we present a brief overview of some of the fractal network models that have been proposed in the literature. Fractal networks are a special class of complex networks that exhibit self-similarity at different scales~\cite{huang2019survey}. Essentially, this means that the structure of the network exhibits similar patterns at both the micro and macro levels, resulting in a complex network topology with a relatively simple generation rule. 

Barabási et al.~\cite{barabasi2001deterministic}, described a deterministic model for creating scale-free networks. The model consists of duplicating two sets of nodes from the previous step and connecting each of the lower nodes of these sets to the root node. A hierarchical rule common to deterministic models is followed, which can be solved analytically. The networks were found to have a high clustering coefficient and a small average path length. Another approach to fractal networks was based on the scale-free network model~\cite{ravasz2003hierarchical} but applied to a modular topology. The resulting networks combine the scale-free property with a high degree of modularity.

Another work describes a different deterministic model for scale-free networks~\cite{jung2002geometric}. The model generates a tree structure in which each node produces a certain number of offspring at each step. The degree distribution follows a power law. The authors also considered loop structures by adding new connection rules. In~\cite{carletti2010weighted} the authors present a construction of weighted fractal networks. It is also shown that weighted fractal networks have non-trivial spectral properties, share some properties with fractal sets, such as self-similarity, and can be used to model some phenomena in physics and biology.
A more recent fractal network model (FSFN) has been described in~\cite{yakubo2022general}. This work considers a small graph called a generator to replace each edge in the previous generation of the network. By selecting different generators, the authors can regulate the structural properties of the network. 

Fractal networks have been widely used to model and analyze various natural and artificial systems~~\cite{kawasaki2010reciprocal}, such as brain networks~\cite{shijo2014topological}, traffic flow~\cite{nagatani2023successive}, communication networks~\cite{kochanski2018flexible}, and quantum transport~\cite{xu2021quantum}. Moreover, fractal networks present interesting properties such as robustness, efficiency, and scalability, as allowed by their branching structure and self-similar nature~\cite{shijo2014topological}. In addition, fractal network models can help researchers understand the properties and behavior of complex systems and design better networks for various other applications~\cite{zheng2016fractal,huang2011improved,wen2021fractal}.

\section{Methodology} \label{sec:methodology}
In this section, we describe the methodology here adopted to analyze and compare artificial networks, obtained from a fractal generator algorithm, with others networks models.

\subsection{Fractal generator}{\label{subsec:model}
The Fractal Generator is an algorithm that has been developed as a means to generate fractal networks based on a predefined small subgraph which shall be henceforth referred to as a \emph{motif}. A network \emph{motif} is a subgraph that appears repeatedly in a given network or in different networks (e.g.~\cite{milo2002network}). The process of constructing a fractal network by successively replacing a pattern can be understood as being analogous to how fractal curves such as Koch's (e.g.~\cite{peitgen2004chaos}) are typically generated.

In this study, we adopted the nine motifs shown in Figure~\ref{fig:motifs}, although the method is not limited to these particular choices. Each motif has a reference node, as indicated by the blue nodes in Figure~\ref{fig:motifs}, that specifies the point of respective insertion into the growing network. Therefore, a same motif can lead to potentially distinct respective networks depending on the choice of reference nodes.

\begin{figure}[!htpb]
  \centering
     \includegraphics[width=0.52\textwidth]{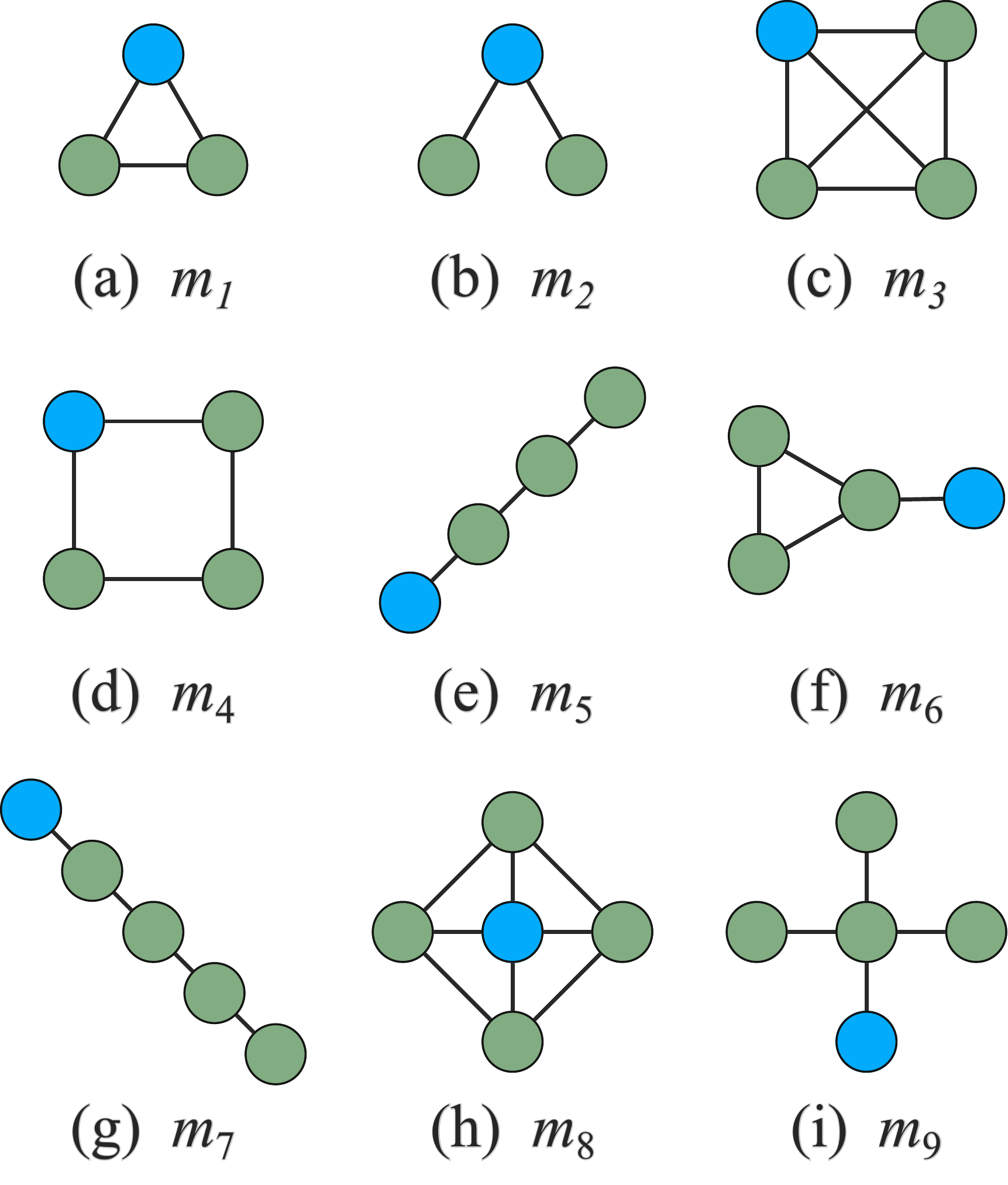}
   \caption{The nine \emph{motifs} adopted in the current work. The reference nodes are shown in blue. In principle, any node of a given motif can be used as a reference node.}
  \label{fig:motifs}
\end{figure}

Figure~\ref{fig:example} shows an example of constructing a fractal network corresponding to the motif shown in Figure~\ref{fig:motifs}(b). Three basic interactions of the growing method are illustrated, starting from interaction 0 which corresponds to the own original motif, as shown in Figure~\ref{fig:example}(a).

\begin{figure}[!htpb]
  \centering
     \includegraphics[width=0.8\textwidth]{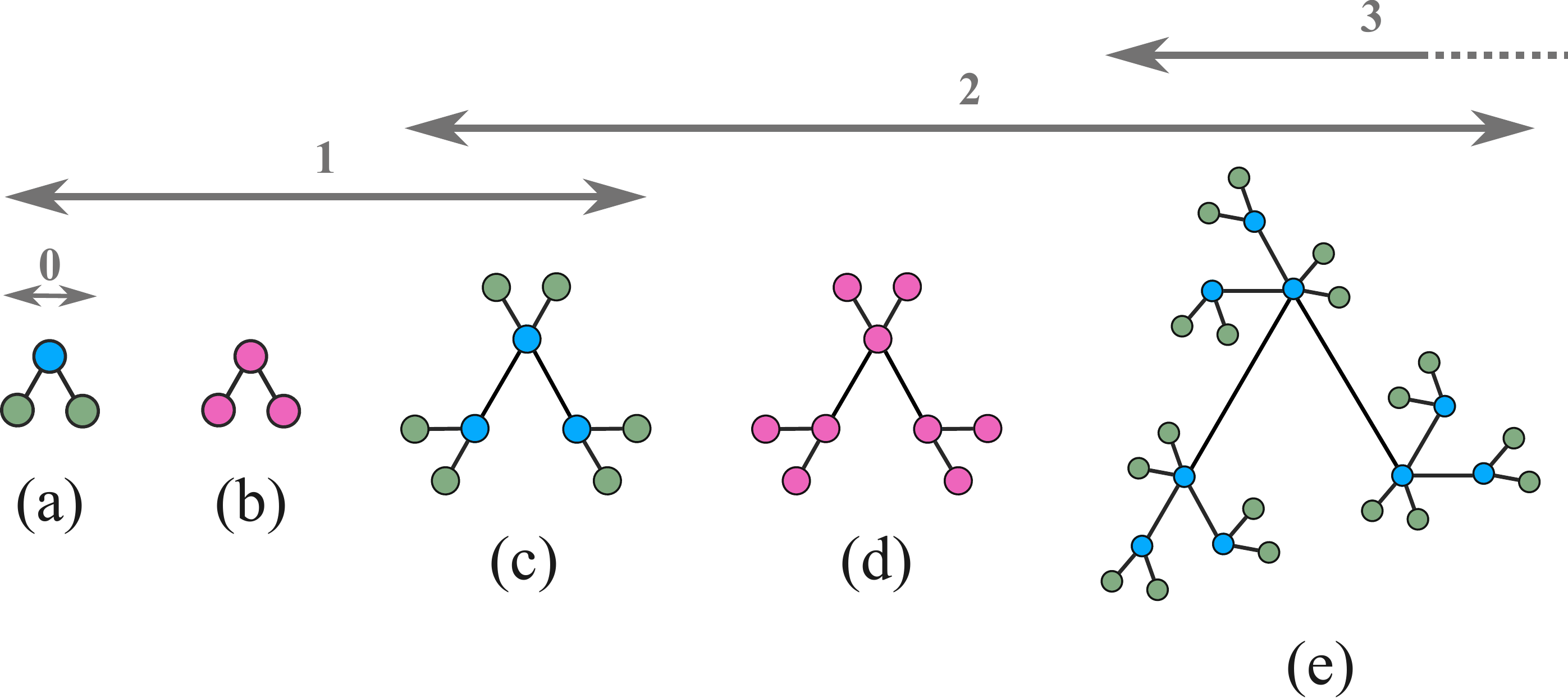}
   \caption{Illustration of the three first interactions in the generation of a fractal network, respectively to a given motif, by using the methodology suggested in the present work. Starting with the chosen motif, each interaction consists of replacing nodes potentialized for change (shown in magenta), by the reference node (marked in blue) of respective instances of the adopted motif. Observe that the positions of the nodes are not specified by the suggested method, which yields only the respective topology. Positions can be obtained, for instance, by considering network visualization algorithms.}
  \label{fig:example}
\end{figure}

In the first interaction, the nodes in the current network to be replaced by the motif (highlighted in magenta in Figure~\ref{fig:example}b), are potentialized for subsequent growth. Each of these nodes is then substituted by the reference node of the corresponding instances of the chosen motif, thereby completing interaction 1. This process is then repeated for subsequent interactions 2, 3, and so on, as shown in Figure~\ref{fig:example}(c)--(e).

The above-described methodology can be applied in different ways. For instance, it is possible to consider only a subset of the nodes of the current network as being potentialized, as illustrated in Figure~\ref{fig:example2}.  In addition, it is also possible to consider more than a single motif, according to several possible manners, while growing heterogeneous networks. 

\begin{figure}[!htpb]
  \centering
     \includegraphics[width=0.8\textwidth]{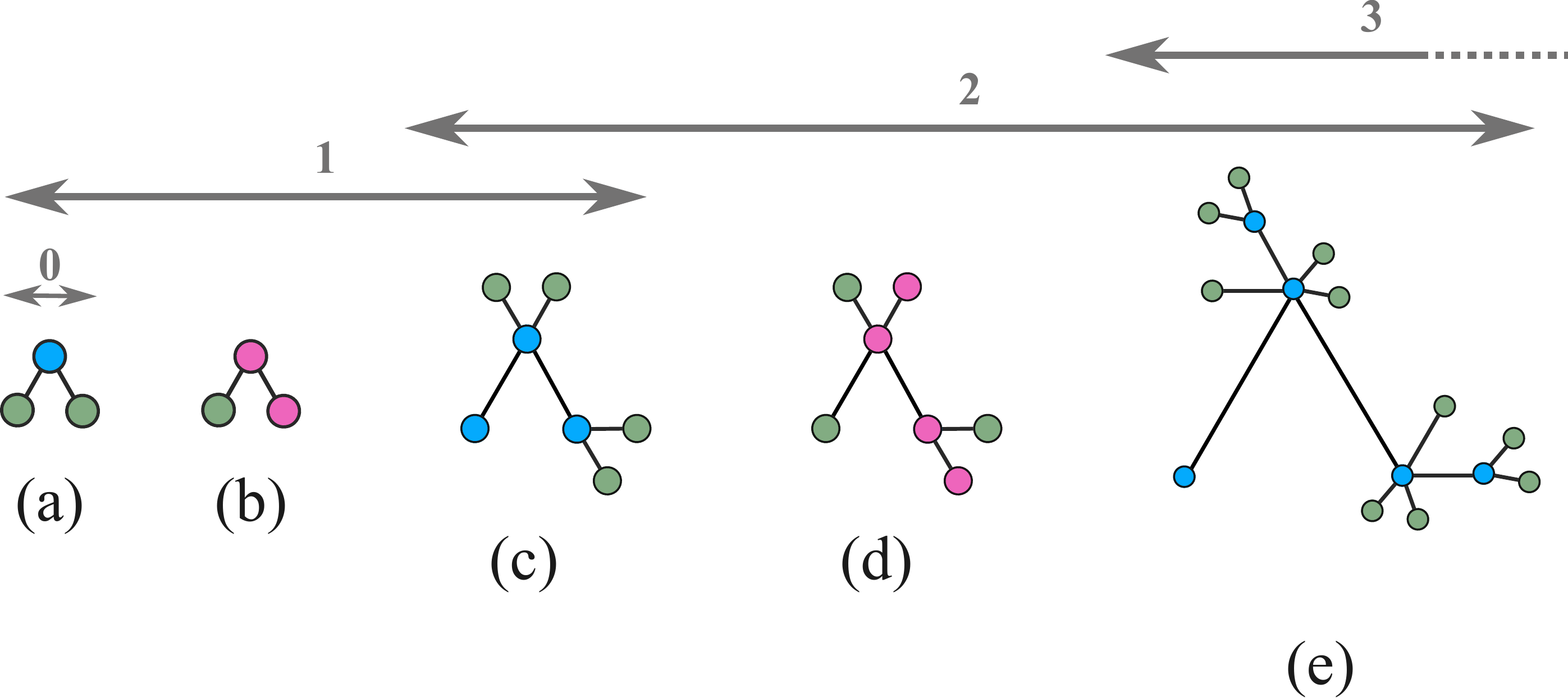}
   \caption{Illustration of the fractal network generation by considering only a deterministic subset of the current network as being potentialized for growth, instead of all respective nodes as adopted in Fig.\ref{fig:example}.}
  \label{fig:example2}
\end{figure}

In the example in Figure~\ref{fig:example2}, the nodes are potentialized in a deterministic manner, with one of the extremity nodes in the original motif being always left without being potentialized. The potentialization can also be implemented randomly. 

An extension of the proposed generation methodology is to consider more than one reference node in the adopted motif, all of them being interconnected with the previous network (or to a subset of respective nodes). Several combinations and variations of the above approaches are also possible.

\subsection{Network measurement}\label{network_measurement}
In order to evaluate the networks generated by the suggested fractal model, we employed some traditional network measurements of distinct topological aspects of a network. The first measurement under consideration was the clustering coefficient~\cite{watts1998collective}, which quantifies how densely connected are the neighbors of a node. For undirected networks, the clustering coefficient of a node $i$ is defined as follows:

\begin{equation}
c_i = \frac{2 l_i}{k_i(k_i - 1)},
\end{equation}
where $k_i$ is the degree of node $i$, and $l_i$ is the number of links among its neighbors.

We also measured the betweenness centrality (e.g.~\cite{freeman1977set}), a widely used measure based on shortest paths. The betweenness centrality of a node $v$ is given by the expression:

\begin{equation}
B(v) = \sum_{ij} \frac{\sigma_{ij}(v)}{\sigma_{ij}} ,
\end{equation}
where $\sigma_{ij}$ is the number of shortest paths that connect $i$ and $j$, and $\sigma_{ij}(v)$ is the number of shortest paths connecting $i$ and $j$ that pass through $v$.

Another measure that has been considered consists in the average shortest path, which returns the shortest path between two given nodes, calculated for all node pairs of a graph. Dijkstra's algorithm~\cite{dijkstra1959note} has been employed to calculate this measure.

The hierarchical degree (e.g.~\cite{da2006hierarchical}), a generalization of node degree, has also been adopted. The hierarchical degree is defined as the number of connections that exist between nodes that are at a distance $d$ and $d-1$ away from a given reference node. Thus, we have the hierarchical degree of order $d$ expressed by the following equation:

\begin{equation}
K_i(d) = \sum_{j \in R^i_{d}} \sum_{l \in R^i_{d-1}} a_{lj} ,
\end{equation}
where $i$ is the reference node, $a_{lj}$ are the elements of the network's adjacency matrix, $R^i_d$ are the sets of nodes at a distance $d$ from $i$, and $R^i_{d-i}$ are the sets of nodes at a distance $d-1$ form $i$.

\subsection{Fractal dimension of networks}

The concept of fractal network~\cite{song2005self} is related to the existence of a power law between the number of boxes ($\emph{Nb}$) and the box size ($\emph{lb}$) used to cover the network, which can be mathematically expressed as:
\begin{equation}
\emph{Nb} \propto \emph{lb}^{-\emph{df}},
\end{equation}
where $\emph{df}$ corresponds to the fractal dimension of the network.

When mapped into a log-log space, the above relationship leads to the respective straight-line relationship:
\begin{equation}
\log(\emph{Nb}) \propto -\emph{df} \ \log(\emph{lb}).
\end{equation}

In principle, the higher the value of the fractal dimension estimated for a given complex network, the more intricate its topology will tend to be, in the sense of presenting more complex branches and cycles structure. Therefore, the fractal dimension provides an interesting resource for quantifying the intricacy  of the topology of specific complex networks.

\section{Results and discussion} \label{sec:results}

In this section, we present the results of applying the proposed method to generate fractal networks, discuss the results, and then analyze the generated networks in terms of respective measurements and fractal dimensions. As a means of evaluating the growth progress of the fractal networks, the measures outlined in Section ~\ref{network_measurement} have been calculated and then projected into a respective PCA (Principal Component Analysis) space ~\cite{jolliffe2011principal, gewers2021principal}.

\subsection{Examples of fractal networks}

In this section, we present examples of fractal networks generated by our proposed model described in Subsection~\ref{subsec:model}.

\begin{figure*}[!htpb]
  \centering
     \includegraphics[width=0.99\textwidth]{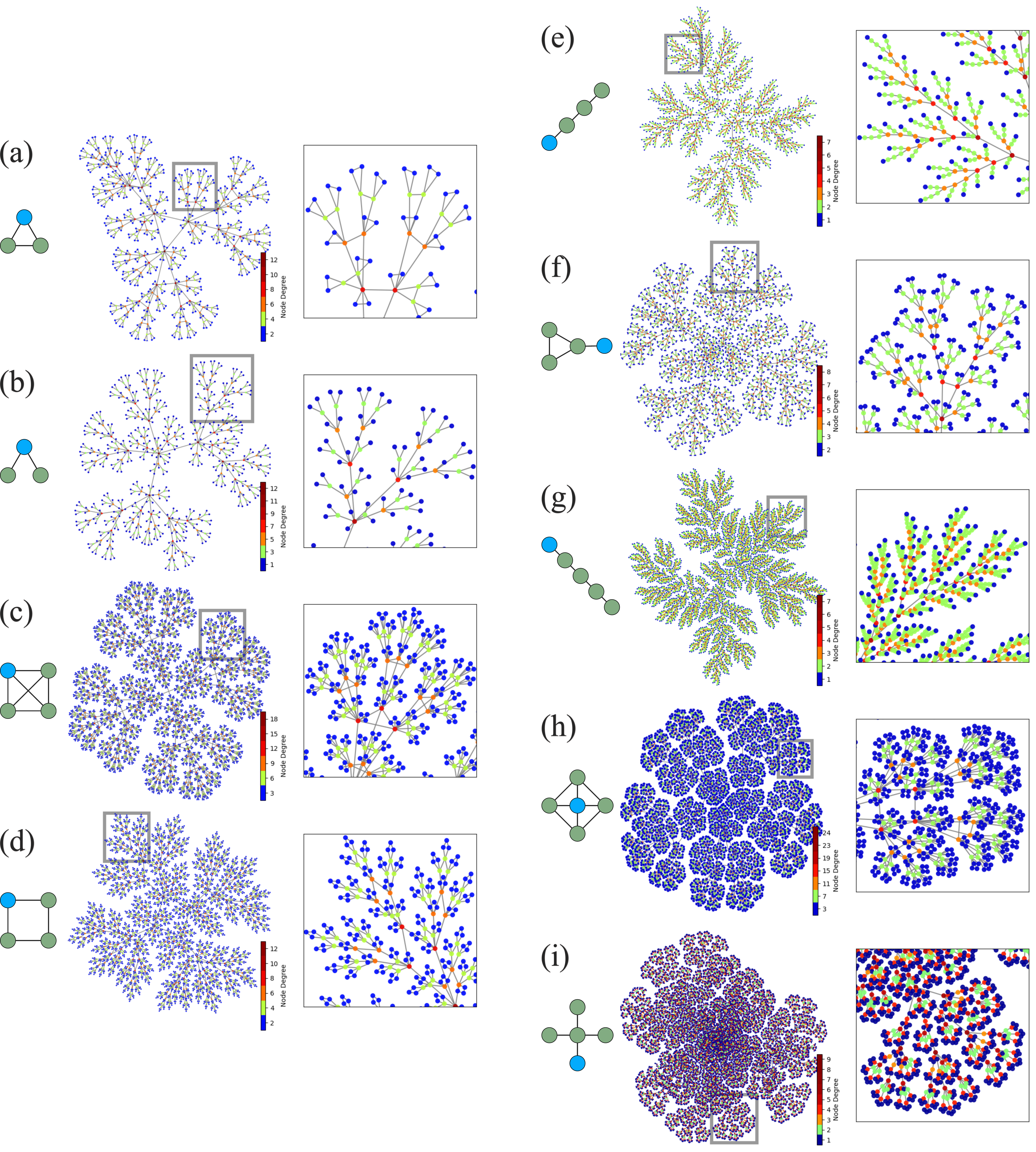}
   \caption{The several complex networks obtained from the nine considered motifs. Also shown are zoomed portions of each respective structure. The color bar represents the node degree. Markedly distinct topologies can be obtained, varying from loop-less trees (b, e, g, i) to more intricate structures including cycles (a, c, d, f, h) and well-defined modules (h).}
  \label{fig:Example_Networks}
\end{figure*}

Figure~\ref{fig:Example_Networks} shows the networks generated from different motifs using the fractal generator algorithm, considering 5 growth stages.

One of the main observations from these examples is that all the networks obtained have a predominantly hierarchical structure, as expected from their definition and growing rule. This means that there are few high-degree nodes and many low-degree nodes, as typically observed in trees.

In addition, we have that the considered motifs and reference nodes, although simple, lead to networks with distinct topologies. In a way, there is an amplification of the original differences between the motifs. For instance, motifs without loops produce networks without loops, as shown in Figures~\ref{fig:Example_Networks} (b), (e), and (g), while motifs that contain loops produce networks containing loops, as shown in Figures~\ref{fig:Example_Networks} (a), (c), (d), (f), (h), and (i).  

Moreover, in some cases, the resulting networks are characterized by a modular structure, where groups of nodes are densely connected to each other and sparsely connected to other groups. This is more evident for highly connected motifs, such as Figure~\ref{fig:Example_Networks} (h).

Figure~\ref{fig:F_example} shows an example of a fractal network generated by using the motif (f). The figure illustrates the first four stages of the network growth, also indicating the number of nodes in each stage. As the network grows, it becomes more intricate and exhibits self-similar patterns at a wider range of scales.

\begin{figure}[!htpb]
  \centering
    \includegraphics[width=0.65\textwidth]{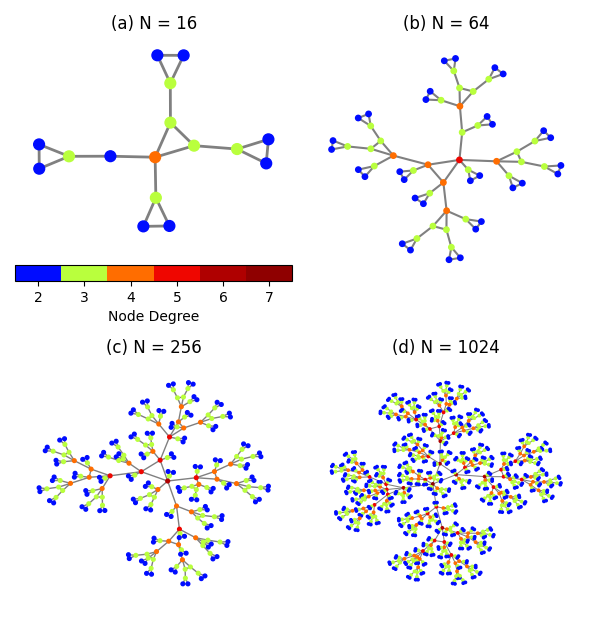}
   \caption{Illustration of the four initial stages of the fractal network, determined by the adopted motif (f) and the corresponding reference node. Observe that the topology of the obtained networks tends to become increasingly more elaborated along these stages.}
  \label{fig:F_example}
\end{figure}

\subsection{Comparison between networks generated by different motifs}

The effect of the initial \emph{motif} and how it can influence the growth dynamics was analyzed by comparing the measurements of the topology of the networks obtained from the \emph{Fractal generator}. 

More specifically, we evaluated the growth progress of networks generated by the proposed fractal model in this work in terms of the four following measurements: clustering coefficient, betweenness centrality, shortest path, and hierarchical degree (order 2). These measurements, described in Section~\ref{network_measurement}, have been calculated as the respective averages over all network nodes.

\begin{figure}[!htpb]
  \centering
    \includegraphics[width=0.7\textwidth]{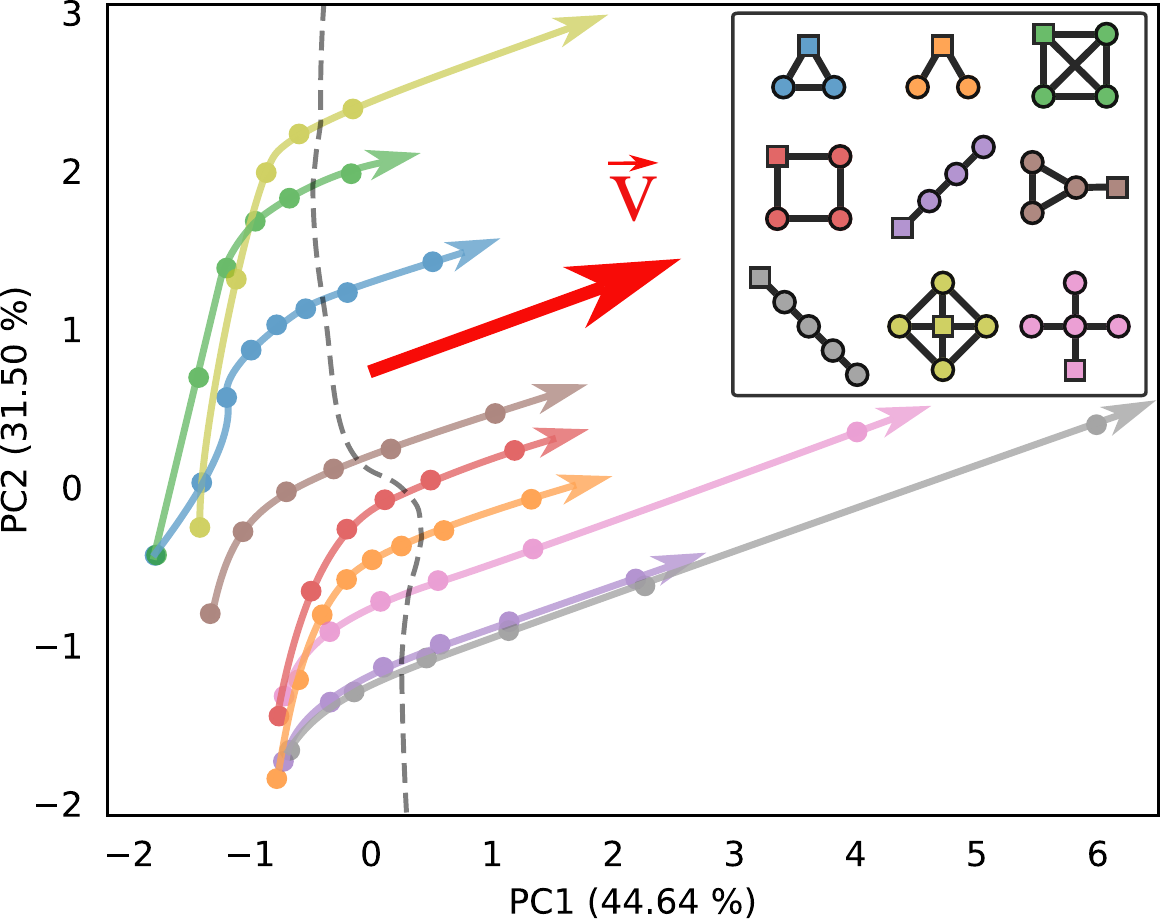}
   \caption{PCA of fractal network measurements. Each color indicates a different \emph{motif}, and the arrows indicate the sense in the increasing number of iterations of the fractal generator. The reference node in each motif has been identified as a square.}
  \label{fig:PCA_motfs}
\end{figure}

We use a PCA to reduce the dimensionality of the measurement space to two dimensions, allowing the visualization of the evolution of the networks generated by the fractal model. Figure~\ref{fig:PCA_motfs} shows the PCA results obtained for the nine motifs used to construct the networks. The percentage of explained variance is shown at the respective axes, indicating that a total of $76.14\%$ of the original data variation has been accounted for by two-dimensional PCA projection. Each color represents a different motif, and each point corresponds to a network obtained by applying the fractal model. The properties of the networks obtained at each stage of their growth can be thought of as defining \emph{trajectories} in their corresponding feature spaces. The arrows indicate the direction of network growth along these trajectories.

Interestingly, two well-defined regions can be readily discerned along each of the obtained trajectories, corresponding to a \emph{transient} portion which is followed by a respective \emph{equilibrium} region. In order to separate these two regions, we calculated the inclination at each of the points along each curve and took the respective portion that exceed the threshold of $0.109$ as corresponding to an estimation of the transient regime.

The figure shows that the distances between the original motifs increase as the networks grow along the transient regime. This implies that the networks diverge from each other and become more distinct as they increase in size. The motifs $m_1$, $m_3$, and $m_8$ show larger trajectory changes than the other six motifs. Interestingly, these three cases correspond to the more intensely connected motifs among the nine motifs considered and are therefore relatively more intricate than the others. This suggests a direct relationship between the initial topology of the motifs and the changes in topological properties during the transient region.

Another interesting point to note is that the divergence reaches a steady state, in most cases, after three or four interactions, indicating a \emph{equilibrium} regime. The dashed line in Figure~\ref{fig:PCA_motfs} separates the transient and equilibrium phases along all trajectories. In the equilibrium phase, the trajectories are mostly parallel, which means that the trajectory velocities have a constant orientation, represented by the red vector $\vec{V}$ in Figure~\ref{fig:PCA_motfs}. This result suggests that in the equilibrium regime, the effect on topological properties is similar among all networks.

To further analyze the changes in the topological properties of all considered networks, we use the reference vector ($\vec{V}$), which is mostly parallel to all trajectories in the equilibrium regime. This vector, which has been normalized to have unit magnitude, has been estimated to be:
\begin{align}
    \vec{V} = \left( 0.000, 0.850, 0.259, 0.017 \right)
\end{align}

The four components of this vector correspond respectively to clustering coefficient, betweenness centrality, shortest path, and hierarchical degree.  Observe that this reference vector accounts only for the \emph{orientation} of the equilibrium trajectories, and not their respective speed magnitudes,  which can be observed from Figure~\ref{fig:PCA_motfs} not to remain constant.

The betweenness centrality of the networks can thus be understood as contributing to most of the changes that the considered networks undergo as they grow along their respective equilibrium regimes. The second largest measure corresponds to the average shortest path length of the networks and therefore also has a relatively important contribution to characterizing the growth of the networks at equilibrium.

The dominant role of the betweenness for the topological changes observed along the trajectories can be better understood by considering that all considered networks correspond mostly to trees, with possible loops being relatively sparse and local.  In a tree network where new branching orders are successively added, the betweenness of all nodes will increase sharply, which is indeed observed in the experimental results.

Since the new orders of motifs added to the considered growing networks tend to be limited to local topological changes, a corresponding progressive increase in the average shortest path length would indeed be observed, which is confirmed by the markedly increasing separation between successive points in the respective trajectories (trajectory speed magnitude).

We used the box-covering method to estimate the fractal dimension ($df$) of each network generated, considering 5 growing stages.  Figure~\ref{fig:dimension} presents the results obtained using the Fuzzy box-covering method~\cite{zhang2014fuzzy, wei2013box} to calculate the fractal dimension, which are shown in the inset.

\begin{figure}[!htpb]
  \centering
    \includegraphics[width=0.7\textwidth]{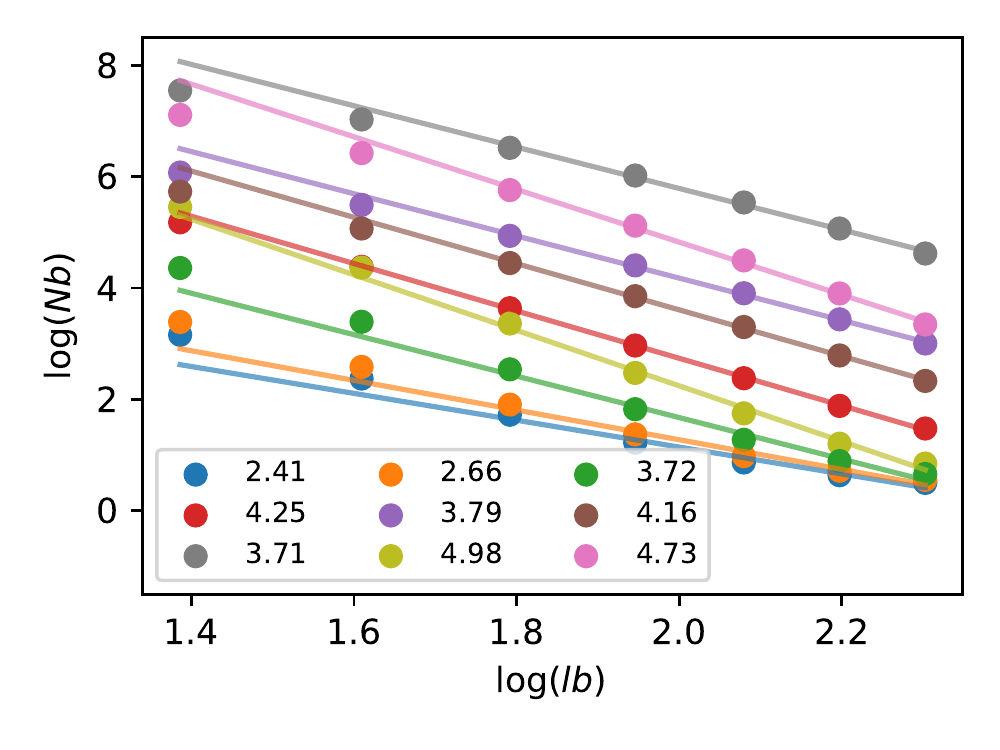}
   \caption{Log-log plots and respectively obtained fractal dimensions (estimated by the box-covering method) of the complex networks generated from the nine considered motifs.}
  \label{fig:dimension}
\end{figure}

Overall, a good adherence can be observed between the log-log curves and the respective least mean square approximations, especially for the larger values of $lb$, in which case the subsequent intervals become smaller. A range of fractal dimensions has been respectively obtained, varying from $2.41$ to $4.98$, which supports that complex networks with accentuated distinct topological intricacy can be respectively obtained by the choice of different motifs.

However, other aspects including the choice of reference nodes and numbers of growth steps, can also potentially influence the resulting topological structure of the obtained networks. Figure~\ref{fig:comparison} depicts the log-log plots and respective fractal dimensions of obtained for networks generated from motif $m6$ by considering all the three possible choices (considering group symmetries) of reference nodes.  

\begin{figure}[!htpb]
  \centering
    \includegraphics[width=0.7\textwidth]{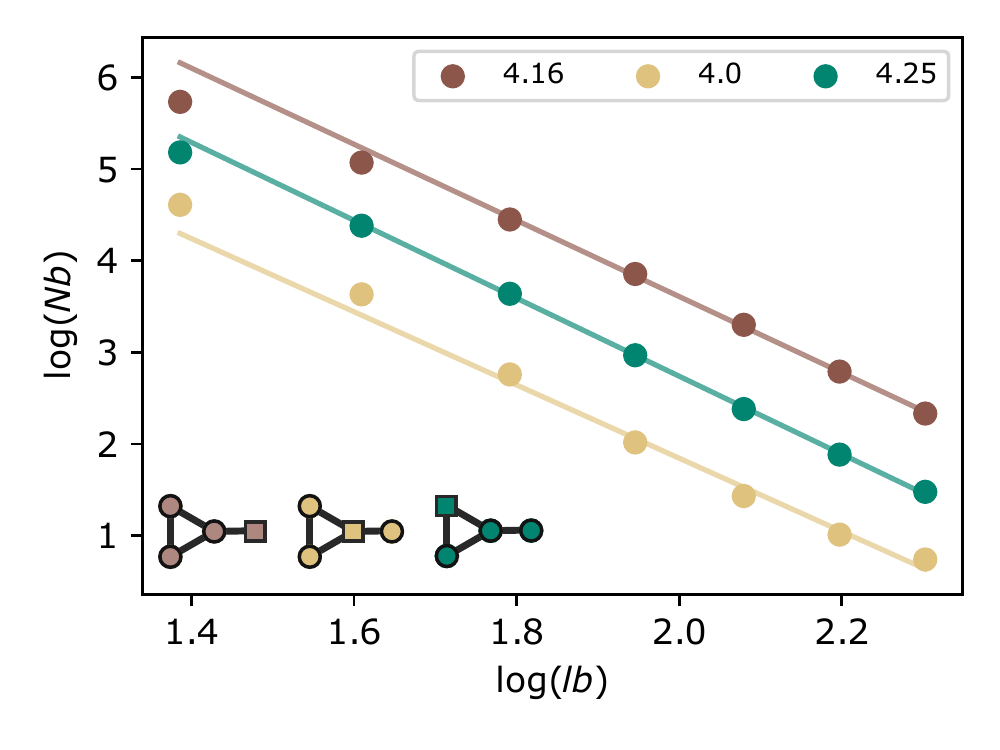}
   \caption{Log-log plots and respectively estimated fractal dimensions of networks obtained from motif $m6$ by taking different choices of nodes as reference (shown as square). This result illustrates the potential influence of the reference node on the resulting topological properties of the respectively generated networks.}
  \label{fig:comparison}
\end{figure}

The results illustrated in Figure~\ref{fig:comparison} confirm that different choices of nodes to be taken as references can greatly impact on the respectively obtained fractal dimension.

\section{Concluding Remarks}

Complex networks have been understood mostly as graphs that exhibit an intricate topology when compared to uniform or regular counterparts. In particular, complex networks cannot be effectively described in terms of the average node degree, or even of a small set of topological measures. A particularly interesting aspect of such networks is how their topological intricacy extends along successive hierarchies. The study of fractal networks is an interesting and promising way to understand hierarchical structures, as they potentially exhibit intrinsic self-similarity across topological scales, in both abstract and real-world settings.

In this work, we introduce a method for creating fractal networks by substituting reference nodes from specific motifs in an existing network. We also present and examine several fractal networks derived from nine different motifs. The visualizations show that particularly intricate networks can emerge from relatively simple motifs, and some of the networks have loops and/or modules, while others correspond to trees. In our analyses, we applied visualization and statistical methods in order to better understand how their respective topological features change along successive growth stages.

In the current study, we examined how the topological features of the considered networks changed along the growth stages. We used a PCA projection to plot the trajectories of these features and identify some tendencies. The growth process was found to include two regimes: transient and equilibrium, which were respectively identified and analyzed. For the transient dynamics, the obtained results indicated that more intricate initial motifs tend to lead to larger changes in the corresponding topological measures. In the equilibrium regime, the trajectories resulted mostly parallel to each other, which allowed us to determine a corresponding parallel vector summarizing all the observed equilibrium portions of the trajectories. The study also found that measures related to betweenness centrality and average shortest path tended to be more significant during the equilibrium regime, allowing a more effective characterization of the topological changes implied by the fractal nature of each network.

The log-log plots of the number of boxes in terms of their respective size revealed a good adherence to a straight line, supporting the fractal properties of the obtained networks. In addition, markedly distinct fractal dimensions have been obtained while considering different motifs and/or reference nodes.

The reported results pave the way for several future research, such as exploring alternative growth methods using multiple reference nodes and motifs or using non-deterministic rules. Another possibility would be to use additional measurements to enhance our understanding of the topological characteristics of the fractal networks.

\section*{Acknowledgments}
Alexandre Benatti thanks Coordenação de Aperfeiçoamento de Pessoal de N\'ivel Superior - Brasil (CAPES) - Finance Code 001 (88882.328749/2019-01). Luciano da F. Costa thanks CNPq (grant no. 307085/2018-0) and FAPESP grant 2015/22308-2. The authors thank Filipi N.~Silva (Indiana University Network Science Institute, USA) for graph visualization resources (\url{https://pypi.org/project/heliosFR/}).

\bibliography{references}
\bibliographystyle{unsrt}

\end{document}